\documentstyle[fleqn,epsfig]{article}
\newcounter{figureno}
\newenvironment{capt}{
\phantom{mmmm}
\vspace*{10mm}
\parindent=0pt
\addtocounter{figureno}{1}

\begin{minipage}[t]{160mm}
\small\sl Figure~\thefigureno.\ }{\end{minipage}
\vspace*{-5mm}}
\begin{document}
%%%%%%%%%%%%%%%%%%%%%%%%%%%%%%%%%%%%%%%%%%%%%%%%%%%%%%%%%%%%%%%%%%%%%%%%
\renewcommand{\textfraction}{0}
\renewcommand{\bottomfraction}{1}
% more macros:
\newcommand{\beq}{\begin{equation}}
\newcommand{\eeq}{\end{equation}}
\newcommand{\beqa}{\begin{eqnarray}}
\newcommand{\eeqa}{\end{eqnarray}}
\newcommand{\nn}{\nonumber}

\newcommand{\dd}{\mbox{{\rm d}}}

\newcommand{\Dis}[1]{$\displaystyle #1$}
\newcommand{\Disp}[1]{{\displaystyle #1}}

\newcommand{\hc}{\mbox{{\rm h.c.}}}
\newcommand{\MNS}{M_2}
\newcommand{\mlN}{M_1}
\newcommand{\Mu}{m_{{ u}}}
\newcommand{\Au}{A_{{ u}}}
\newcommand{\Md}{m_{{ d}}}
\newcommand{\Ad}{A_{{ d}}}
\newcommand{\Ab}{A_{{ b}}}
\newcommand{\At}{A_{{ t}}}
\newcommand{\suq}{\tilde{{ u}}}
\newcommand{\Tsp}{\mbox{\scriptsize T}}
\newcommand{\eps}{\epsilon}
\newcommand{\sdq}{\tilde{{ d}}}
\newcommand{\gluino}{\tilde g}
\newcommand{\squark}{\tilde q}
\newcommand{\thW}{\theta_{\rm W}}
\newcommand{\mW}{m_{{ W}}}
\newcommand{\MsQU}{\wtilde{3}{0.8}{M}_{\hspace*{-1mm}U}}
\newcommand{\MsQT}{\wtilde{3}{0.8}{M}_{\hspace*{-1mm}T}}
\newcommand{\Msu}{\wtilde{3}{0.2}{m}_{U}}
\newcommand{\Msd}{\wtilde{3}{0.2}{m}_{\!D}}
\newcommand{\Msq}{m_{\tilde{q}}}
\newcommand{\MsT}{\wtilde{3}{0.2}{m}_{T}}
\newcommand{\Ms}{\wtilde{3}{0.8}{m}}
\newcommand{\wtilde}[3]{\settowidth{\ltT}{\Dis{#3}}
\makebox[\ltT]{$\rule{#2\mmh}{0mm}
\widetilde{\makebox[#1\mm]{\Dis{#3\rule{#2\mm}{0mm}}}}$}}

\newlength{\ltT}
\newlength{\mmh}
\setlength{\mmh}{0.5mm}
\newlength{\mm}
\setlength{\mm}{1mm}

\def\Im{\mbox{\rm Im\ }}
\def\Re{\mbox{\rm Re\ }}
\def\fourth{\textstyle{1\over4}}
\def\gsim{\mathrel{\rlap{\raise 1.5pt \hbox{$>$}}\lower 3.5pt
\hbox{$\sim$}}}
\def\lsim{\mathrel{\rlap{\raise 1.5pt \hbox{$<$}}\lower 3.5pt
\hbox{$\sim$}}}
\def\GeV{{\rm GeV}}
\def\TeV{{\rm TeV}}
\def\Order{{\cal O}}
%%%%%%%%%%%%%%%%%%%%%%%%%%%%%%%%%%%%%%%%%%%%%%%%%%%%%%%%%%%%%%%%%%%%%%%%
\catcode`@=11
\def\citer{\@ifnextchar [{\@tempswatrue\@citexr}{\@tempswafalse\@citexr[]}}

% \citer as abbreviation for 'citerange' replaces the ',' by a '--'
%

\def\@citexr[#1]#2{\if@filesw\immediate\write\@auxout{\string\citation{#2}}\fi
  \def\@citea{}\@cite{\@for\@citeb:=#2\do
    {\@citea\def\@citea{--\penalty\@m}\@ifundefined
       {b@\@citeb}{{\bf ?}\@warning
       {Citation `\@citeb' on page \thepage \space undefined}}%
\hbox{\csname b@\@citeb\endcsname}}}{#1}}
\catcode`@=12
%%%%%%%%%%%%%%%%%%%%%%%%%%%%%%%%%%%%%%%%%%%%%%%%%%%%%%%%%%%%%%%%%%%%%%%%
%
\def\slash#1{#1 \hskip -0.5em /}
%
%%%%%%%%%%%%%%%%%%%%%%%%%%%%%%%%%%%%%%%%%%%%%%%%%%%%%%%%%%%%%%%%%%%%%%%%%%
\def\Month{\ifcase\month\or
January\or February\or March\or April\or May\or June\or 
July\or August\or September\or October\or November\or December\fi}
\def\slash#1{#1 \hskip -0.5em /}
%%%%%%%%%%%%%%%%%%%%%%%%%%%%%%%%%%%%%%%%%%%%%%%%%%%%%%%%%%%%%%%%%%%%%%%%
%
% Here we go:

\begin{flushright}
{\tt
\hfill
University of Bergen, Department of Physics \\
Scientific/Technical Report No.\ 1996-01 \\ ISSN~0803-2696\\
hep-ph/9601284 \\
%\Month, \the\year
January, 1996
}
\end{flushright}
\vspace*{1cm}

%%%%%%%%%%%%%%%%%%%%%%%%%%%%%%%%%%%%%%%%%%%%%%%%%%%%%%%%%%%%%%%%%%%%%%%%
%                          Title
\begin{center}
{\Large \bf The Two-Photon Decay Mode of MSSM Higgs Bosons\\
at the Large Hadron Collider}
\footnote{To appear in:
{\it Proceedings of Xth International Workshop: High Energy
Physics and Quantum Field Theory}, Zveni\-gorod, Russia,
September 20--26, 1995}

%                      author/address
\vspace{4mm}
Bjarte Kileng \\
NORDITA, Blegdamsvej 17, DK-2100 Copenhagen \O, Denmark\\
\vspace{3mm}
Per Osland \\
University of Bergen, All\'egt.~55, N-5007 Bergen, Norway\\
\vspace{3mm}
and \\
\vspace{3mm}
P.N. Pandita\footnote{Permanent address:
North Eastern Hill University, Laitumkhrah, Shillong 793003, India.}\\
Universit\"at Kaiserslautern, Fachbereich Physik,
Erwin - Schr\"odinger - Strasse, \\
D-67663 Kaiserslautern, Germany
\end{center}

%%%%%%%%%%%%%%%%%%%%%%%%%%%%%%%%%%%%%%%%%%%%%%%%%%%%%%%%%%%%%%%%%%%%%%%%
%                       Abstract
%%%%%%%%%%%%%%%%%%%%%%%%%%%%%%%%%%%%%%%%%%%%%%%%%%%%%%%%%%%%%%%%%%%%%%%%
\begin{abstract}
At the Large Hadron Collider (LHC),
the CP-even Higgs bosons ($h^0$ and $H^0$)
of the Minimal Supersymmetric Standard Model (MSSM) 
will be searched for mainly through their two-photon decay.
We present a detailed analysis of
the production and two-photon decay of the CP-even Higgs bosons of MSSM
at the LHC by taking into account all the parameters of the model, especially
the bilinear parameter $\mu$ and the trilinear
supersymmetry breaking parameter $A$.
Non-zero values of these parameters lead to significant mixing
in the squark sector, and, thus, affect the masses of Higgs bosons
through radiative corrections,
as well as their couplings to squarks.
The dependence of the cross section for the production of Higgs,
and its subsequent decay
to two photons,
on various parameters of the MSSM is described in detail.
The cross section times the two-photon branching ratio 
of $h^0$ is of the order of 15--25~fb in much
of the parameter space that remains after imposing the
present  experimental constraints on the parameters.
For the $H^0$, the two-photon branching ratio is only significant
if it is light. With a light $H^0$ the cross section times the
branching ratio may be 200~fb or more.
\end{abstract}

%%%%%%%%%%%%%%%%%%%%%%%%%%%%%%%%%%%%%%%%%%%%%%%%%%%%%%%%%%%%%%%%%%%%%%%%
\section{Introduction}
\label{sec:intro}
\setcounter{equation}{0}
%%%%%%%%%%%%%%%%%%%%%%%%%%%%%%%%%%%%%%%%%%%%%%%%%%%%%%%%%%%%%%%%%%%%%%%%
In the Minimal Supersymmetric Standard Model (MSSM)
\cite{MSSM}, two Higgs doublets ($H_1$ and $H_2$) with opposite 
hypercharge are
required in order to give masses to up--and down--quarks (leptons), 
and to cancel gauge anomalies. 
The physical Higgs boson spectrum in the MSSM consists of 
two $CP$-even neutral bosons $h^0$ and $H^0$, a $CP$-odd neutral
boson $A^0$ and a pair of charged Higgs bosons $H^\pm$.
The most important production mechanism for the neutral SUSY Higgs 
bosons at the Large Hadron Collider (LHC) is the gluon
fusion mechanism, $pp\to gg\to h^0$, $H^0$, $A^0$
\cite{GGMN},
and the Higgs radiation off top and bottom quarks 
\cite{Kunsztetal}.
Except for the small range in the parameter space where the
heavy neutral Higgs $H^0$ decays into a pair of $Z$ bosons,
the rare $\gamma\gamma$ decay mode, apart from $\tau\tau$ decays,
is a promising mode to detect the neutral Higgs particles,
since $b$ quarks are hard to separate from the QCD background. 
It has been pointed out
\cite{KunsztZ,Baer}
that the lightest Higgs 
could be detected in this mode for sufficiently large values 
of the mass of the pseudoscalar Higgs boson $m_A\gg m_Z$.
The $\gamma\gamma$ channel is also important 
for the discovery of $H^0$ for $50~\GeV\le m_A\le 150~\GeV$.

Here we present results of a recent study 
\cite{KOP} of the hadronic production
and subsequent two-photon decay of the $CP$-even Higgs bosons
($h^0$ and $H^0$) of the MSSM,
which is valid for the LHC energy of $\sqrt{s}=14~\TeV$, 
using gluon distribution functions based on recent HERA
data \cite{Plothow},
in order to reassess the feasibility of observing the $CP$-even Higgs
bosons in this mode.
After the completion of this work, a related study was presented
by Kane et al.\ \cite{Kane}.
As mentioned earlier, the gluon fusion mechanism is the dominant
production mechanism of SUSY Higgs bosons in high-energy
$pp$ collisions throughout the entire Higgs mass range.
We study the cross section for the production of the $h^0$
and $H^0$, and their decays, taking into account all 
the parameters of the Supersymmetric Standard Model.
In particular, we take into account
the mixing in the squark sector, the chiral mixing.
This also affects the Higgs boson masses 
through appreciable radiative corrections,
and was previously shown to lead to large corrections to the rates
\cite{Kileng}. 

In the calculation of the production of the Higgs through
gluon-gluon fusion, we include in the triangle graph
all the squarks, as well as $b$ and $t$ quarks, the lightest quarks
having a negligible coupling to the $h^0$.
On the other hand, in the calculation of decay of the Higgs to two
photons, we include in addition to the above, all the sleptons,
$W^\pm$, charginos and the charged Higgs boson.

An important role is played in our analysis by the bilinear
Higgs coupling $\mu$, which occurs in the Lagrangian through the term
\begin{equation}
{\cal L}
= \biggl[
    - \mu \hat{H}^{\Tsp}_{1} \eps \hat{H}_{2}\biggr]_{\theta\,\theta}
    + \hc,
\end{equation}
where $\hat{H}_{1}$ and $\hat{H}_{2}$ are the Higgs superfields with
hypercharge $-1$ and $+1$, respectively.
Furthermore, the Minimal Supersymmetric Model contains several 
soft supersymmetry-breaking terms. 
We write the relevant soft terms in the Lagrangian 
as follows \cite{GirGri}
\begin{eqnarray}
\label{EQU:Lagrangefour}
{\cal L}_{\mbox{{\scriptsize Soft}}}
& = & \Biggl\{ \frac{g\Md\Ad}{\sqrt{2}\;\mW\cos\beta}
      Q^{\Tsp}\eps H_{1}\sdq^{R}
    - \frac{g\Mu\Au}{\sqrt{2}\;\mW\sin\beta}
      Q^{\Tsp}\eps H_{2}\suq^{R}
    + \hc \Biggr\} \nonumber \\
& & - \MsQU^{2}Q^{\dagger}Q - \Msu^{2}\suq^{R\dagger}\suq^{R}
    - \Msd^{2}\sdq^{R\dagger}\sdq^{R}
    - M_{\!H_{1}}^{2} H_{1}^{\dagger}H_{1}
    - M_{\!H_{2}}^{2} H_{2}^{\dagger}H_{2} \nn \\*
& & + \frac{\mlN}{2}\left\{\lambda\lambda +\bar{\lambda}\;
      \bar{\lambda}\right\}
    + \frac{\MNS}{2} \sum_{k=1}^{3} \left\{\Lambda^{k}\Lambda^{k}
    + \bar{\Lambda}^{k}\bar{\Lambda}^{k}\right\},
\end{eqnarray}
with subscripts $u$ (or $U$) and $d$ (or $D$) referring to up and
down-type quarks.
The Higgs production cross
section and the two-photon decay rate depend significantly 
on several of these parameters.
We shall in particular focus on the dependence on the trilinear
couplings $A_d$ and $A_u$.

The Higgs production cross section and the two-photon decay rate 
depend on the gaugino and squark masses, the latter being determined
by, apart from the soft-supersymmetry breaking trilinear coefficients
($A_u$, $A_d$) and the Higgsino mixing parameter $\mu$, the soft
supersymmetry-breaking masses, denoted in 
eq.~(\ref{EQU:Lagrangefour})
by $\MsQU$, $\Msu$ and $\Msd$, respectively.
For simplicity, we shall consider the situation where
$\MsQU=\Msu=\Msd\equiv\Ms$, with $\Ms$ chosen to be
150~GeV for the first two generations, and varied over the values 
150, 500 and 1000~GeV for the third generation.

The contributions of the squark and Higgs bosons to the decay rate 
depend on the relative sign between the parameters $A$ and $\mu$, 
but not on their overall signs.
We note that the Higgs sector depends on $A$ and $\mu$ through
radiative corrections.
The chargino contribution is independent of 
$A$, but depends on the relative sign between $\mu$ and $\MNS$. 
Thus, the $h^{0}\to\gamma\gamma$ decay rate 
is independent of the the overall signs of $A$, $\mu$ and $\MNS$, 
but depends on all the relative signs of these  parameters. 
In most of the parameter space, however,  the dependence on
the chargino mass (and therefore on the sign of $\MNS$) is
rather weak.  In these regions it suffices to consider $A$ positive
and vary the sign of $\mu$.
Finally, the signs of the off-diagonal terms in the squark mass matrices
are determined  by the definition of $A$ and $\mu$, and also
by the definition of the fermion masses \cite{KOP}.

In Sec.~2,  we shall study the implications of the nonzero 
values of $A$ and $\mu$ on the Higgs masses, together with the
constraints related to the other relevant masses. We shall then go on 
to study the cross sections and decay rates for the lighter and the 
heavier CP-even Higgs bosons in Secs.~3 and 4, respectively. 

%%%%%%%%%%%%%%%%%%%%%%%%%%%%%%%%%%%%%%%%%%%%%%%%%%%%%%%%%%%%%%%%%%%%%%%%
\section{The Parameter Space}
\label{sec:limit}
%\setcounter{equation}{0}
%%%%%%%%%%%%%%%%%%%%%%%%%%%%%%%%%%%%%%%%%%%%%%%%%%%%%%%%%%%%%%%%%%%%%%%%
In this section we describe in detail the parameter space relevant for 
the production and decay of the lightest Higgs boson at LHC, and the 
theoretical and experimental constraints on it before presenting
cross sections and decay rates.

At the tree level, the masses of the $CP$-even neutral Higgs 
bosons are given by ($m_{h^0}\le m_{H^0}$) \cite{HHG}
\begin{equation}
m^2_{H^0,h^0}=\frac{1}{2}\left[m_A^2+m_Z^2
\pm\sqrt{\left(m_A^2+m_Z^2\right)^2-4m_Z^2 m_A^2\cos^22\beta}\right],
\end{equation}
which are controlled by two parameters, $m_A$ (the mass of 
the $CP$-odd Higgs boson) and 
$\tan\beta$ ($=v_2/v_1$, where $v_2$ and $v_1$ are the vacuum
expectation values of the two Higgs doublets).
Indeed, the entire Higgs sector at the tree level can be described 
in terms of these two parameters alone.
The corresponding eigenstates are
\begin{eqnarray}
H^0&=&\sqrt{2}\left[(\Re H_1^0-v_1)\cos\alpha 
                   +(\Re H_2^0-v_2)\sin\alpha\right], \\
h^0&=&\sqrt{2}\left[-(\Re H_1^0-v_1)\sin\alpha 
                    +(\Re H_2^0-v_2)\cos\alpha\right],
\end{eqnarray}
where $H_1^0$ and $H_2^0$ are the neutral components of the two
Higgs doublets $H_1$ and $H_2$, respectively.
The $CP$-even mixing angle $\alpha$ is defined through
\beq
\cos2\alpha
=-\cos2\beta\left(\frac{m_{A}^2-m_Z^2}{m_{H^0}^2-m_{h^0}^2}\right),
\qquad -{\pi\over2}\le\alpha\le0.
\eeq

There are, however, substantial radiative corrections 
\cite{Okada}
to the $CP$-even neutral Higgs masses. In the one-loop effective potential
approximation, the radiatively corrected squared mass matrix for the
CP-even Higgs bosons has the following form:
\begin{equation}
\label{eq:M-matrix}
M^2=\left(\matrix{\sin^2\beta\, m_A^2+\cos^2\beta\, m_Z^2 &
                  -\sin\beta\cos\beta(m_A^2+m_Z^2) \cr
                  -\sin\beta\cos\beta(m_A^2+m_Z^2) &
                  \cos^2\beta\, m_A^2+\sin^2\beta\, m_Z^2 }\right)
+\left(\matrix{\Delta_{11} & \Delta_{12} \cr
               \Delta_{21} & \Delta_{22}}\right),
\end{equation}
where the radiative corrections $\Delta_{ij}$ depend on, 
besides the top quark mass, 
the bilinear parameter $\mu$ in the superpotential, 
the soft supersymmetry breaking trilinear couplings ($\Au$, $\Ad$), 
and masses ($\MsQU$, $\Msu$, $\Msd$, etc.), as well as $\tan\beta$.
In the limit of no mixing, $\mu=\Au=\Ad=0$, so that
$\Delta_{11}=\Delta_{12}=\Delta_{21}=0$, with
\begin{equation}
\label{eq:Deltas}
\Delta_{22}=\frac{3g^2 m_t^4}{8\pi^2 m_W^2\sin^2\beta}
\log\left(1+\frac{\Ms^2}{m_t^2}\right), \qquad
\mu=A_u=A_d=0.
\end{equation}
The radiative corrections are, in general, positive, and they shift
the mass of the lightest neutral Higgs boson upwards \cite{Okada}.
More recent radiative corrections to the Higgs sector \cite{Carena}
which are valid when the squark masses are
of the same order of magnitude, 
have not been taken into account in our study \cite{KOP}.
As long as the ``loop particles" are far from threshold for
real production, the cross section does not depend very strongly
on the exact value of the Higgs mass.

In order to simplify the calculations,
we shall assume that all the trilinear couplings are equal so that
\beq
\Au=\Ad\equiv A.
\eeq
Furthermore, we shall take the top-quark mass to be 176~GeV
\cite{CDF} in our numerical calculations.
The parameters that enter the neutral $CP$-even Higgs mass
matrix are varied in the following ranges:
\beqa
50~\GeV&\leq& m_A \leq 1000~\GeV, \qquad 1.1\leq\tan\beta\leq 50.0,
\nn \\
\qquad 50~\GeV&\leq&|\mu|\leq1000~\GeV, \qquad
0\leq A\leq 1000~\GeV.
\eeqa
These values cover essentially the entire physically interesting
range of parameters in the MSSM.
However, not all of the above parameter values are allowed
because of the experimental constraints on the squark, chargino
and $h^0$ masses.
For low values of $\Ms$, the lightest squark tends to be
too light (below the most rigorous experimental bound of
$\sim 44.5~\GeV$ \cite{squarklimit}), or even unphysical
(mass squared negative).
%%%%%%%%%%%%%%%%%%%%%%%%%%%%%%%%%%%%%%%%%%%%%%%%%%%%%%%%%%%%%%%%%%%%%%%%
\begin{figure}[htb]
\begin{center}
\setlength{\unitlength}{1cm}
\begin{picture}(16,8.0)
\put(0.,-2.0)
{\mbox{\epsfysize=10.5cm\epsffile{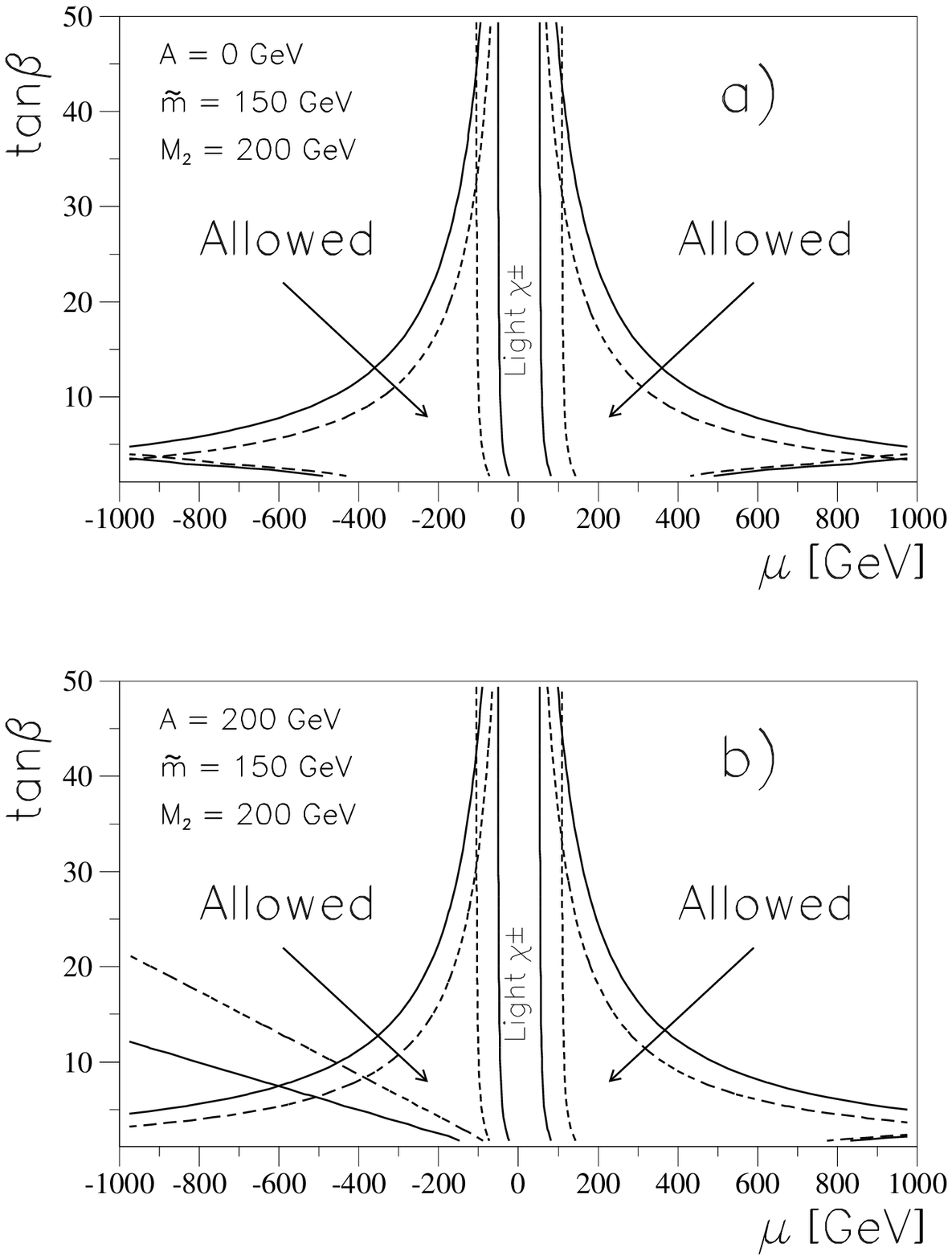}}\hspace*{-5mm}
 \mbox{\epsfysize=10.5cm\epsffile{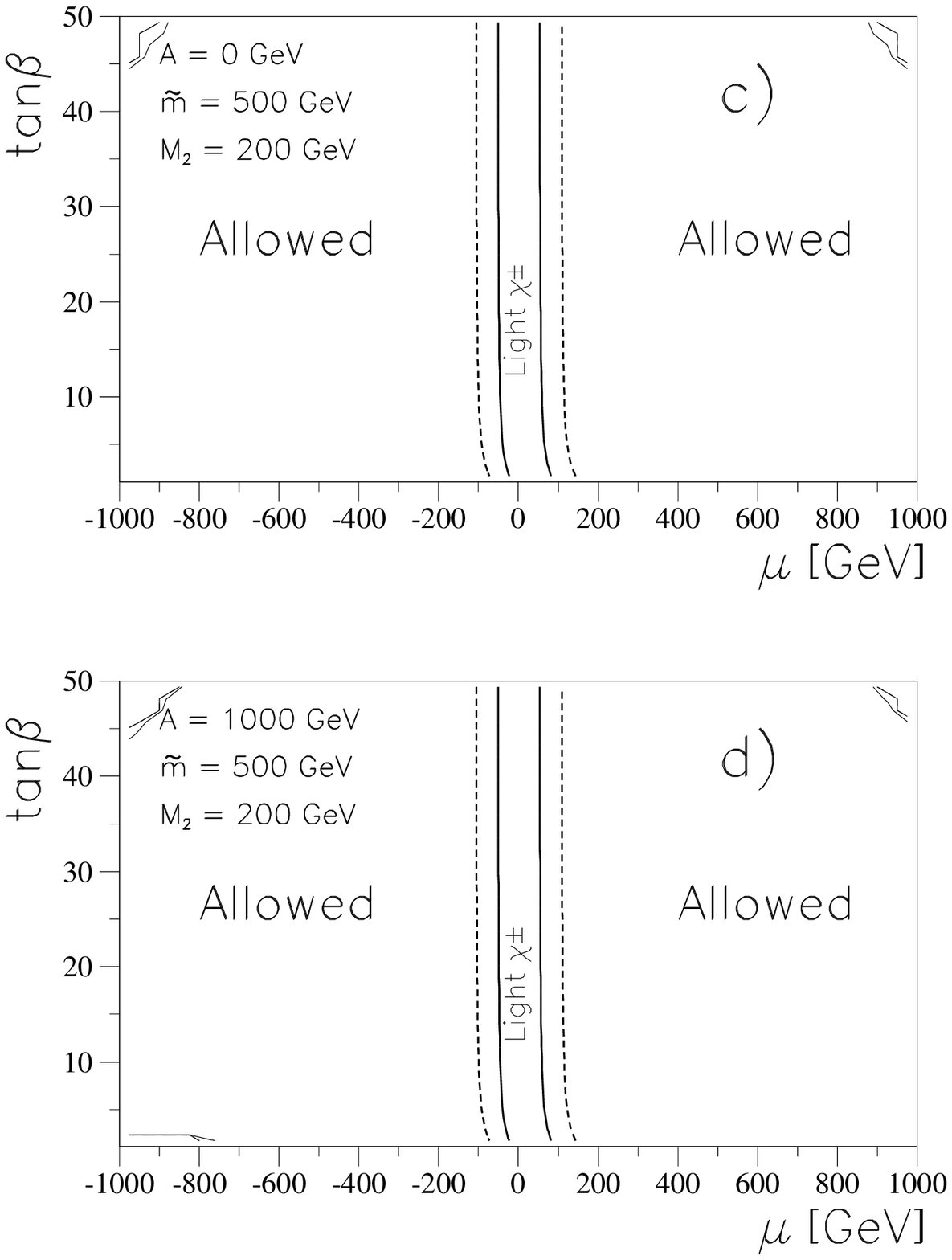}}}  
\end{picture}
\vspace*{10mm}
\begin{capt}
Regions in the $\mu$-$\tan\beta$ plane which are ruled out
by too light chargino ($\chi^\pm$), squark and Higgs masses.
The gaugino mass scale is $\MNS=200~\GeV$ and $m_A=200~\GeV$.
The solid (dashed) contours for small $|\mu|$ refer to the chargino mass
$m_{\chi^\pm}=45$ (90)~GeV.
Two values of $\Ms$ are considered, {\em left}: $\Ms=150~\GeV$,
{\em right}: $\Ms=500~\GeV$.
Two values of the trilinear mixing parameter $A=0$ are considered:
a) and c) $A=0$, b) $A=200~\GeV$, and d) $A=1000~\GeV$.
For $\Ms=150~\GeV$, the squark masses are too light or
unphysical in much of the $\mu$-$\tan\beta$ plane.
The hyperbola-like contours give regions that are excluded
by the lightest $b$ squark being below 45~GeV (solid) or
90~GeV (dashed).
The more straight contours at large $\mu$ and small $\tan\beta$ 
similarly indicate regions that are excluded by the
lightest $t$ squark.
For $\Ms=500~\GeV$ the unlabelled contours near the corners at large 
$|\mu|$ refer to regions where the $h^0$ mass would be below 45~GeV.
\end{capt}
\end{center}
\end{figure}
%%%%%%%%%%%%%%%%%%%%%%%%%%%%%%%%%%%%%%%%%%%%%%%%%%%%%%%%%%%%%%%%%%%%%%%%
The excluded region of the parameter space is shown
in the left part of fig.~1 for $\Ms=150~\GeV$, $\MNS=200~\GeV$, 
$m_A=200~\GeV$ and for two values of the trilinear coupling $A$.
The allowed region in the $\mu-\tan\beta$ plane decreases 
with increasing $A$, but the dependence on $\MNS$ and $m_A$ 
in this region is rather weak.
In order to have acceptable $b$-squark masses,
$\mu$ and $\tan\beta$ must lie {\it inside} of the
hyperbola-shaped curves. Similarly, in order to have
acceptable $t$-squarks, the corners at large $|\mu|$ and
small $\tan\beta$ must be excluded.
%%%%%%%%%%%%%%%%%%%%%%%%%%%%%%%%%%%%%%%%%%%%%%%%%%%%%%%

%\medskip\noindent{\bf The chargino}

The chargino masses are, at the tree level, given by the expression
\beqa
m^2_{\chi^\pm}
&=& \frac{1}{2}(\MNS^2+\mu^2)+\mW^2 \nn \\
& & \pm\left[\frac{1}{4}(\MNS^2-\mu^2)^2 +\mW^4\cos^2{2\beta}
+\mW^2(\MNS^2+\mu^2+2\mu\MNS\sin{2\beta})\right]^{1/2}.
\eeqa
For $\mu=0$, we have
\beq
m^2_{\chi^\pm} = \frac{1}{2}\MNS^2+\mW^2
\pm\left[\frac{1}{4}\MNS^4+\mW^2\cos^2{2\beta}+\mW^2\MNS^2\right]^{1/2}.
\eeq
When $\mu=0$, we see that, for $\tan\beta\gg1$,
the lightest chargino becomes massless.
Actually, small values of $\mu$ are unacceptable
for all values of $\tan\beta$.
The lowest acceptable value for $|\mu|$ will depend
on $\tan\beta$, but that dependence is rather weak.
The excluded region due to the chargino being too light,
increases with decreasing values of $\MNS$.
We note that the radiative corrections to the chargino masses are small
for most of the parameter space \cite{Lahanas}.
In fig.~1 we show the contours in the $\mu$-$\tan\beta$
plane outside of which the chargino has an acceptable mass 
($>45~\GeV$) \cite{chino}.
By the time the LHC starts
operating, one would have searched for charginos with
masses up to 90~GeV at LEP2. Contours relevant for LEP2
are also shown.

%\medskip\noindent{\bf The $h^0$}

For larger values of $\Ms$ (right part of fig.~1), 
there is no problem with the squark masses.
However, for large values of $\Ms$ the experimental constraints
on the $h^0$ mass rule out some of the regions of parameter space.
This is illustrated in fig.~1 for $\Ms=500~\GeV$.
The corners at large values of $|\mu|$ and $\tan\beta$ must
be excluded since one of the squarks there would be unphysical 
or the $h^0$ mass would be lower than
the experimental bound obtained at LEP \cite{LEPdata}.
The extent of these forbidden regions in the parameter space
grow rapidly as $m_A$ decreases below $\Order(150~\GeV)$.
They also increase with increasing values of $A$.

As discussed above,
the mass of the lighter $CP$-even Higgs boson $h^0$ will depend
significantly on $\mu$, $\tan\beta$, $A$ and $\Ms$, through
the radiative corrections. For $\Ms=500~\GeV$, and two values
each of $m_A$ (100 and 200~GeV) and $A$ (0 and 1000~GeV),
the dependence on $\mu$ and $\tan\beta$ is displayed in fig.~2.
As already indicated in fig.~1, at large $|\mu|$ and large $\tan\beta$,
the radiative corrections are large and negative, driving the value of
$m_{h^0}$ well {\it below} the tree-level value.
%%%%%%%%%%%%%%%%%%%%%%%%%%%%%%%%%%%%%%%%%%%%%%%%%%%%%%%%%%%%%%%%%%%%%%%%
\begin{figure}[thb]
\begin{center}
\setlength{\unitlength}{1cm}
\begin{picture}(16,10.0)
\put(3.3,-2)
{\mbox{\epsfysize=12cm\epsffile{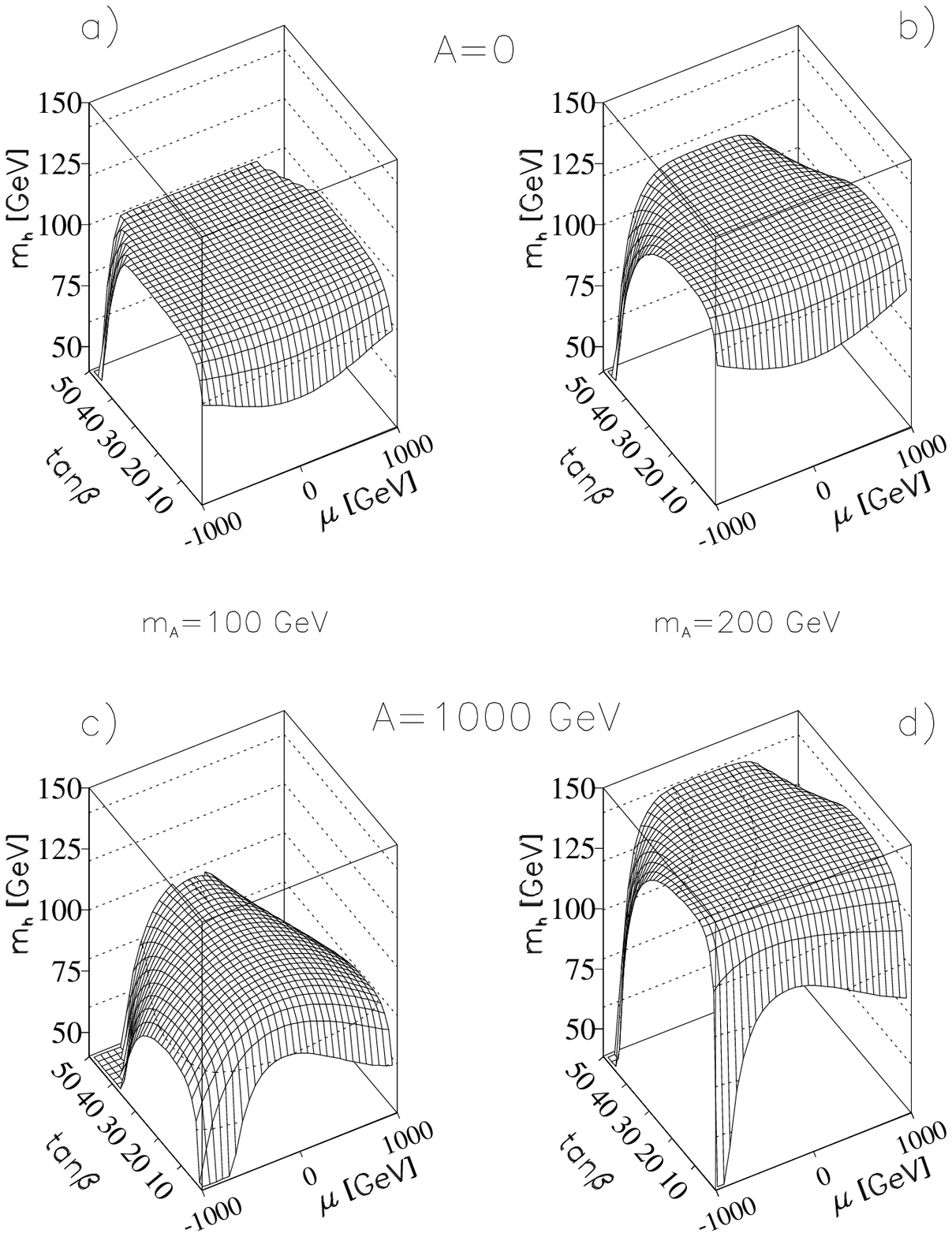}}}
\end{picture}
\begin{capt}
Mass of the lightest $CP$-even Higgs boson vs.\ $\mu$
and $\tan\beta$, 
for $\MNS=200~\GeV$ and $\Ms=500~\GeV$.
Two values of $m_A$ and two values of $A$ are considered: 
a)~$m_A=100~\GeV$, $A=0$,
b)~$m_A=200~\GeV$, $A=0~\GeV$, 
c)~$m_A=100~\GeV$, $A=1000~\GeV$,
d)~$m_A=200~\GeV$, $A=1000~\GeV$.
\end{capt}
\end{center}
%\vspace*{-5mm}
\end{figure}
%%%%%%%%%%%%%%%%%%%%%%%%%%%%%%%%%%%%%%%%%%%%%%%%%%%%%%%%%%%%%%%%%%%%%%%%

%\medskip\noindent{\bf The charged Higgs mass}

The charged Higgs boson mass is given by
\beq
m^2_{H^\pm}=\mW^2+m_A^2+\Delta,
\eeq
where $\Delta$ arises due to radiative corrections
and is a complicated function of the parameters of the model
\cite{Brignole}.

The radiative corrections to the charged Higgs mass are, in general,
not as large as in the case of neutral Higgs bosons.
This is due to an approximate global $SU(2)\times SU(2)$ symmetry
\cite{Haber}, valid in the limit of no mixing.
In certain regions of parameter space the radiative corrections can,
however, be large.
This is the case when the trilinear mixing parameter $A$ is
large, $m_A$ is small, and when furthermore $\tan\beta$ is large.
We shall include the effects of non-zero $A$ and $\mu$
in the calculation of the charged Higgs mass.
The present experimental lower bound of 40--45~GeV
\cite{chargedH} on the charged Higgs is not restrictive,
but presumably by the time the LHC starts operating, one will 
have searched for charged Higgs bosons at LEP2 with mass up to 
around 90~GeV.
Even this bound does not appreciably restrict the parameter space.
                    
%\medskip\noindent{\bf The neutralino}

The neutralino mass matrix depends on the four parameters
$\MNS$, $\mlN$, $\mu$ and $\tan\beta$.
However, one may reduce the number of parameters by assuming that 
the MSSM is embedded in a grand unified theory so that the
SUSY-breaking gaugino masses are equal to
a common mass at the grand unified scale.
At the electroweak scale, we then have \cite{Inoue}
\beq
\mlN=\frac{5}{3}\tan^2\thW\, \MNS.
\eeq
We shall assume this relation throughout in what follows.
The neutralino masses enter the calculation through the total
width of the Higgs boson.
We have considered values of the gaugino mass parameter
$\MNS$ to be 50, 200, or 1000~GeV \cite{KOP}.
We shall here present the numerical results for the case of
$\MNS=200~\GeV$.
The experimental constraint on the lightest
neutralino mass rules out certain regions of the parameter space
\cite{neutr}, but these depend on several parameters, 
and are therefore not reproduced in fig.~1.
They are generally correlated with the bounds on chargino masses
\cite{chino}.
%%%%%%%%%%%%%%%%%%%%%%%%%%%%%%%%%%%%%%%%%%%%%%%%%%%%%%%%%%%%%%%%%%%%%%%%
\section{The lighter $CP$-even Higgs boson $h^0$}
\label{sec:Xsects-h}
%\setcounter{equation}{0}
%%%%%%%%%%%%%%%%%%%%%%%%%%%%%%%%%%%%%%%%%%%%%%%%%%%%%%%%%%%%%%%%%%%%%%%%

The cross section for
\beq
pp\to h^0 X,
\eeq
%%%%%%%%%%%%%%%%%%%%%%%%%%%%%%%%%%%%%%%%%%%%%%%%%%%%%%%%%%%%%%%%%%%%%%%%
\begin{figure}[bht]
\begin{center}
\setlength{\unitlength}{1cm}
\begin{picture}(16,10.0)
\put(3.3,-2)
{\mbox{\epsfysize=12cm\epsffile{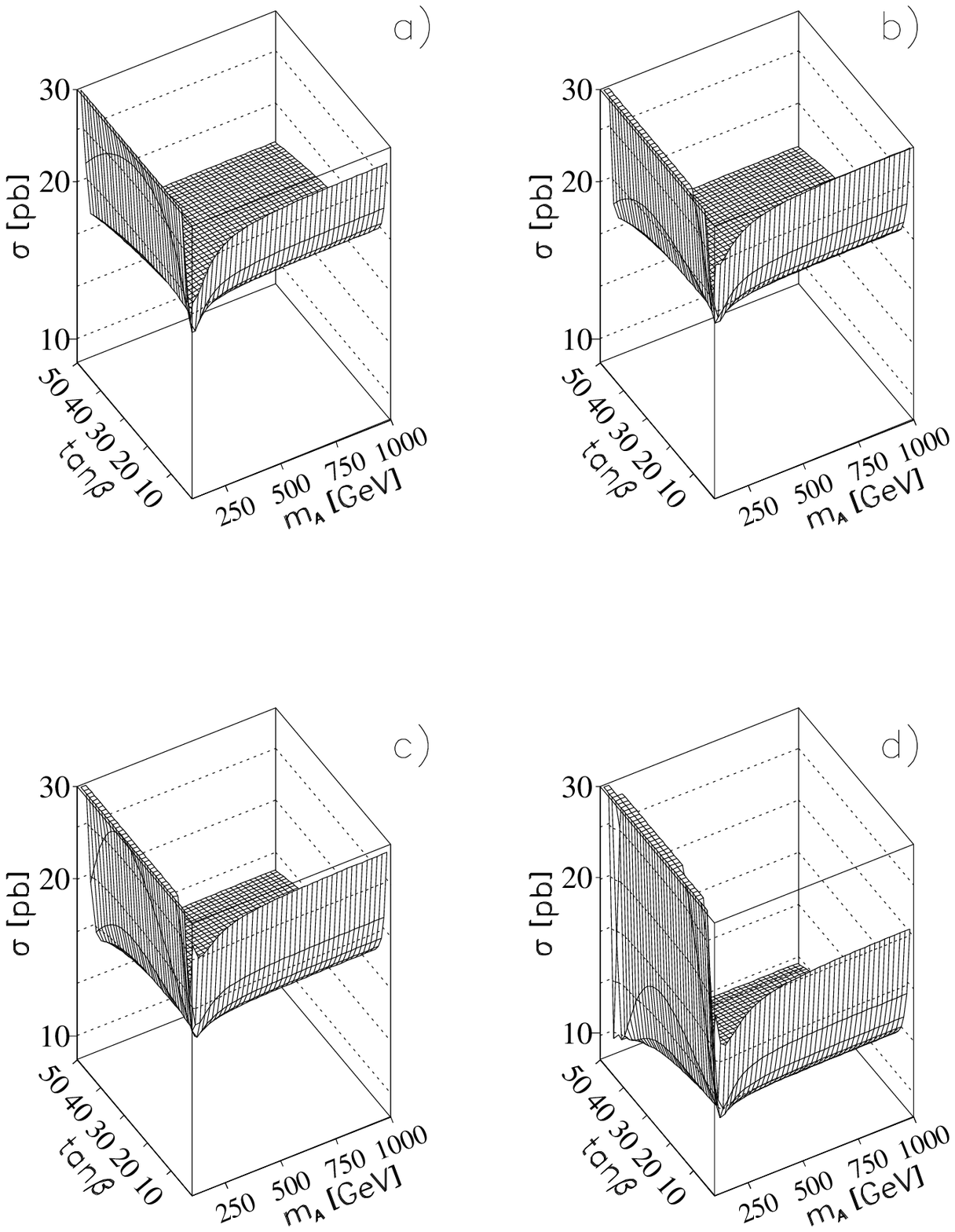}}} 
\end{picture}
\begin{capt}
Cross section for $pp\to h^0X$
as a function of $m_A$ and $\tan\beta$
for $\MNS=200~\GeV$, 
$\Ms=500~\GeV$, and $\mu=500~\GeV$.
Four values of $A$ are considered: a)~$A=0$,
b)~$A=200~\GeV$, c)~$A=500~\GeV$ and d)~$A=1000~\GeV$.
\end{capt}
\end{center}
%\vspace*{-5mm}
\end{figure}
%%%%%%%%%%%%%%%%%%%%%%%%%%%%%%%%%%%%%%%%%%%%%%%%%%%%%%%%%%%%%%%%%%%%%%%%
is calculated from the triangle diagram convoluted with
the gluon distribution functions,
\begin{equation}
\sigma
=\sqrt2\,\pi\,G_{\rm F}\,
\biggl(\frac{\alpha_s}{8\pi}\biggr)^2 \, \frac{m_{h^0}^2}{s}
\Big|\sum_{k} I_k(\tau)\Big|^2 
\int_{-\log(\sqrt s/m_{h^0})}^{\log(\sqrt s/m_{h^0})} {\rm d} y \,
G\Bigl(\frac{m_{h^0}}{\sqrt s}\,e^y\Bigr)\,
G\Bigl(\frac{m_{h^0}}{\sqrt s}\,e^{-y}\Bigr),
\end{equation}
with contributions from various diagrams $k$. For the standard 
case of a top-quark loop,
\begin{equation}
I(\tau)=\frac{\tau}{2}\biggl\{1-(\tau-1)
\biggl[{\rm arcsin}\biggl(\frac{1}{\sqrt \tau}\biggr)\biggr]^2\biggr\},
\nonumber
\end{equation}
and $\tau=(2m_t/m_{h^0})^2>1$.

For $\MNS=200~\GeV$, $\Ms=500~\GeV$, and $\mu=500~\GeV$, 
we show in fig.~3 this cross section for four values of $A$, 
the trilinear coupling parameter.
From this figure the following features are noteworthy:
\begin{itemize}
\item The cross section decreases appreciably for large values of $A$.
This is mainly due to an increase in the $h^0$ mass.
\item The cross section increases sharply for small values of
$\tan\beta$, and also at small $m_A$.
The increase at small $\tan\beta$ is caused by the $h^0$ becoming light.
At small values of $m_A$ and large $A$, the couplings of $h^0$
to $b$ quarks and $\tau$ leptons become large, making
the cross section very large in this region.
\end{itemize}

For the same parameters as above, we show in fig.~4
the total decay rate, $\Gamma(h^0\to\mbox{all})$ and the
two-photon decay rate, $\Gamma(h^0\to\gamma\gamma)$.
%%%%%%%%%%%%%%%%%%%%%%%%%%%%%%%%%%%%%%%%%%%%%%%%%%%%%%%%%%%%%%%%%%%%%%%%
\begin{figure}[thb]
\begin{center}
\setlength{\unitlength}{1cm}
\begin{picture}(16,10.0)
\put(3.3,-2)
{\mbox{\epsfysize=12cm\epsffile{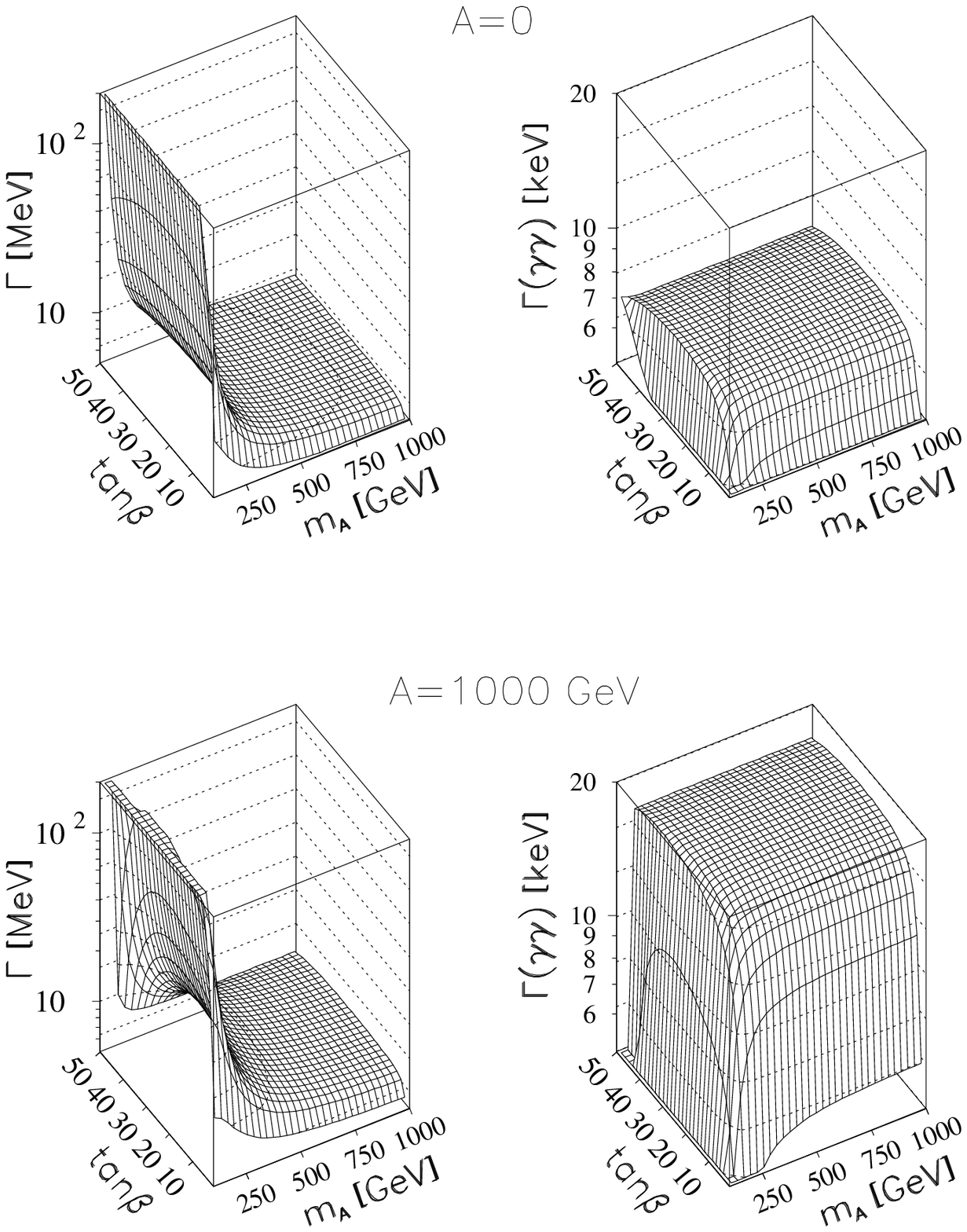}}} 
\end{picture}
\begin{capt}
Total decay rate $\Gamma(h^0\to\mbox{all})$ and
two-photon decay rate $\Gamma(h^0\to\gamma\gamma)$,
as functions  of $m_A$ and $\tan\beta$
for $\MNS=200~\GeV$,
$\Ms=500~\GeV$, and $\mu=500~\GeV$.
Two values of $A$ are considered: $A=0$
and $A=1000~\GeV$.
\end{capt}
\end{center}
\end{figure}
%%%%%%%%%%%%%%%%%%%%%%%%%%%%%%%%%%%%%%%%%%%%%%%%%%%%%%%%%%%%%%%%%%%%%%%%
In contrast to fig.~3, here we only consider two values of $A$,
namely $A=0$ and $A=1000~\GeV$.
The two-photon decay rate is seen to increase sharply at large
values of $A$, but this does not result in a correspondingly larger 
rate for the process
\beq
\label{eq:pp-h0-gammagamma}
pp\to h^0 X \to \gamma\gamma X,
\eeq
since the production cross section also decreases, as shown in fig.~3
(mostly due to an increase in the Higgs mass, $m_{h^0}$).
%%%%%%%%%%%%%%%%%%%%%%%%%%%%%%%%%%%%%%%%%%%%%%%%%%%%%%%%%%%%%%%%%%%%%%%%
\begin{figure}[thb]
\begin{center}
\setlength{\unitlength}{1cm}
\begin{picture}(16,5.5)
\put(0.0,-1)
{\mbox{\epsfysize=6.0cm\epsffile{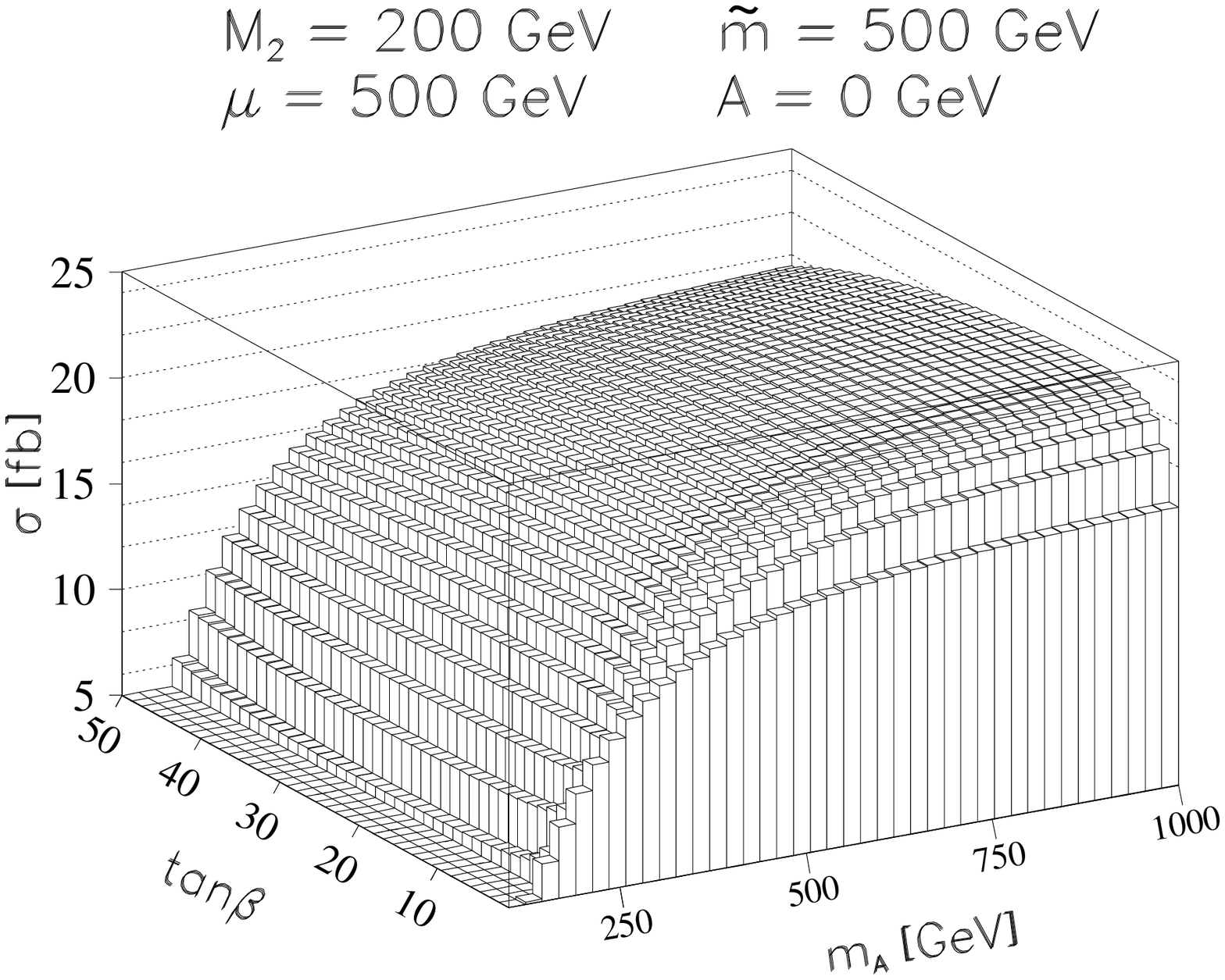}} %=eps_crossgg51.eps
 \mbox{\epsfysize=6.0cm\epsffile{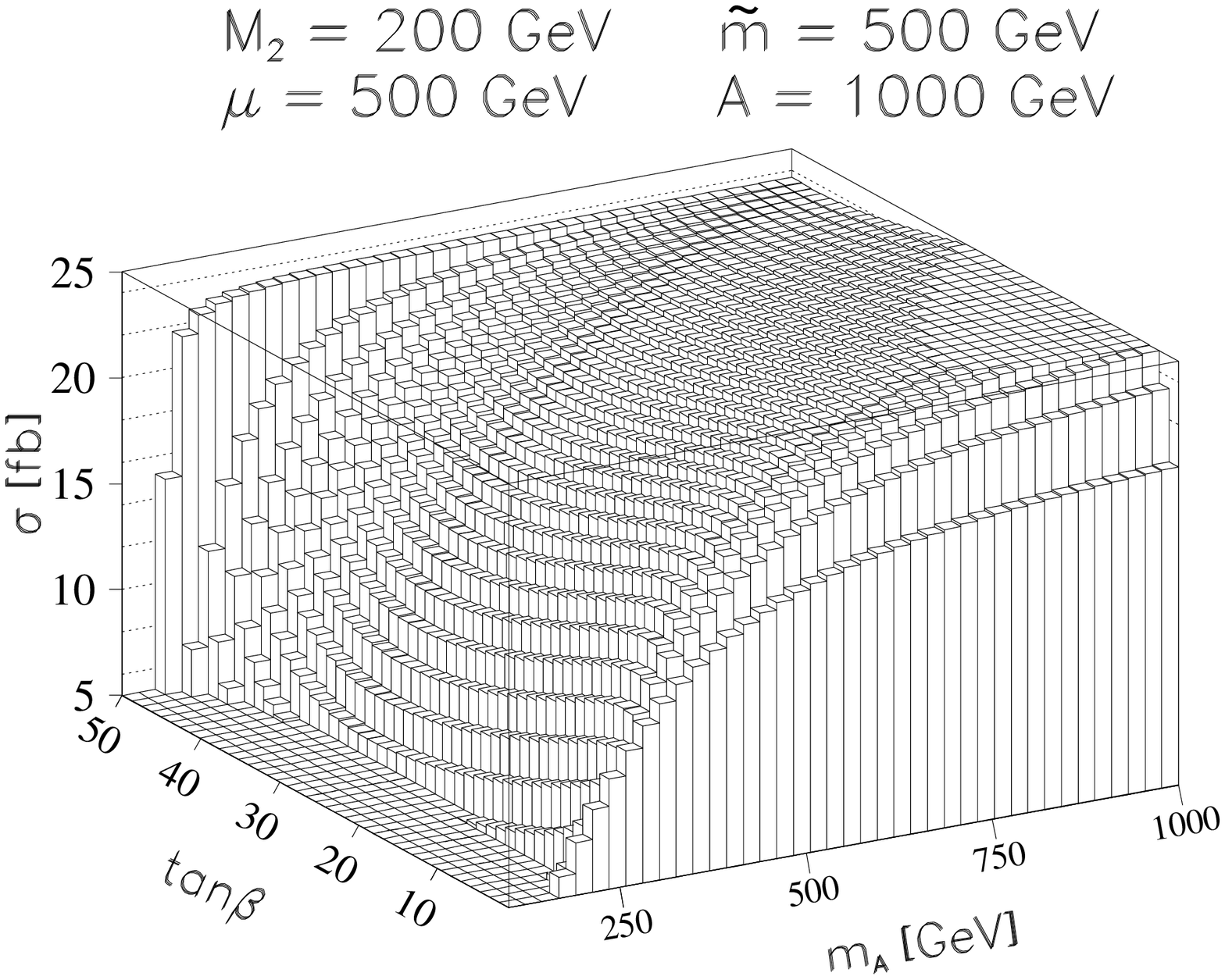}}}
\end{picture}
\begin{capt}
Cross section for $pp\to h^0X\to\gamma\gamma X$ 
as a function of $m_A$ and $\tan\beta$
for $\MNS=200~\GeV$,
$\Ms=500~\GeV$, $\mu=500~\GeV$, and two values of $A$:
{\em left}: $A=0$, {\em right}: $A=1000~\GeV$.
\end{capt}
\end{center}
\end{figure}
%%%%%%%%%%%%%%%%%%%%%%%%%%%%%%%%%%%%%%%%%%%%%%%%%%%%%%%%%%%%%%%%%%%%%%%%
In fig.~5 we show the cross section for the process
(\ref{eq:pp-h0-gammagamma}).
A characteristic feature of the cross section is that it is small
at moderate values of $m_A$, and then increases steadily with
increasing $m_A$, reaching asymptotically a plateau.
This behaviour is caused by the contribution of the $W$ to the
triangle graph for $h^0\to\gamma\gamma$.
The $h^0WW$ coupling is proportional to $\sin(\beta-\alpha)$,
where\footnote{In actual calculations we take the radiatively
corrected formula for $\alpha$.}
\beq
\cos^2(\beta-\alpha)
=\frac{m_{h^0}^2 (m_Z^2 - m_{h^0}^2)}
      {m_{A^0}^2 (m_{H^0}^2 - m_{h^0}^2)}.
\eeq   

%%%%%%%%%%%%%%%%%%%%%%%%%%%%%%%%%%%%%%%%%%%%%%%%%%%%%%%%%%%%%%%%%%%%%%%%
\begin{figure}[thb]
\begin{center}
\setlength{\unitlength}{1cm}
\begin{picture}(16,5.5)
\put(0.0,-1)
{\mbox{\epsfysize=6.0cm\epsffile{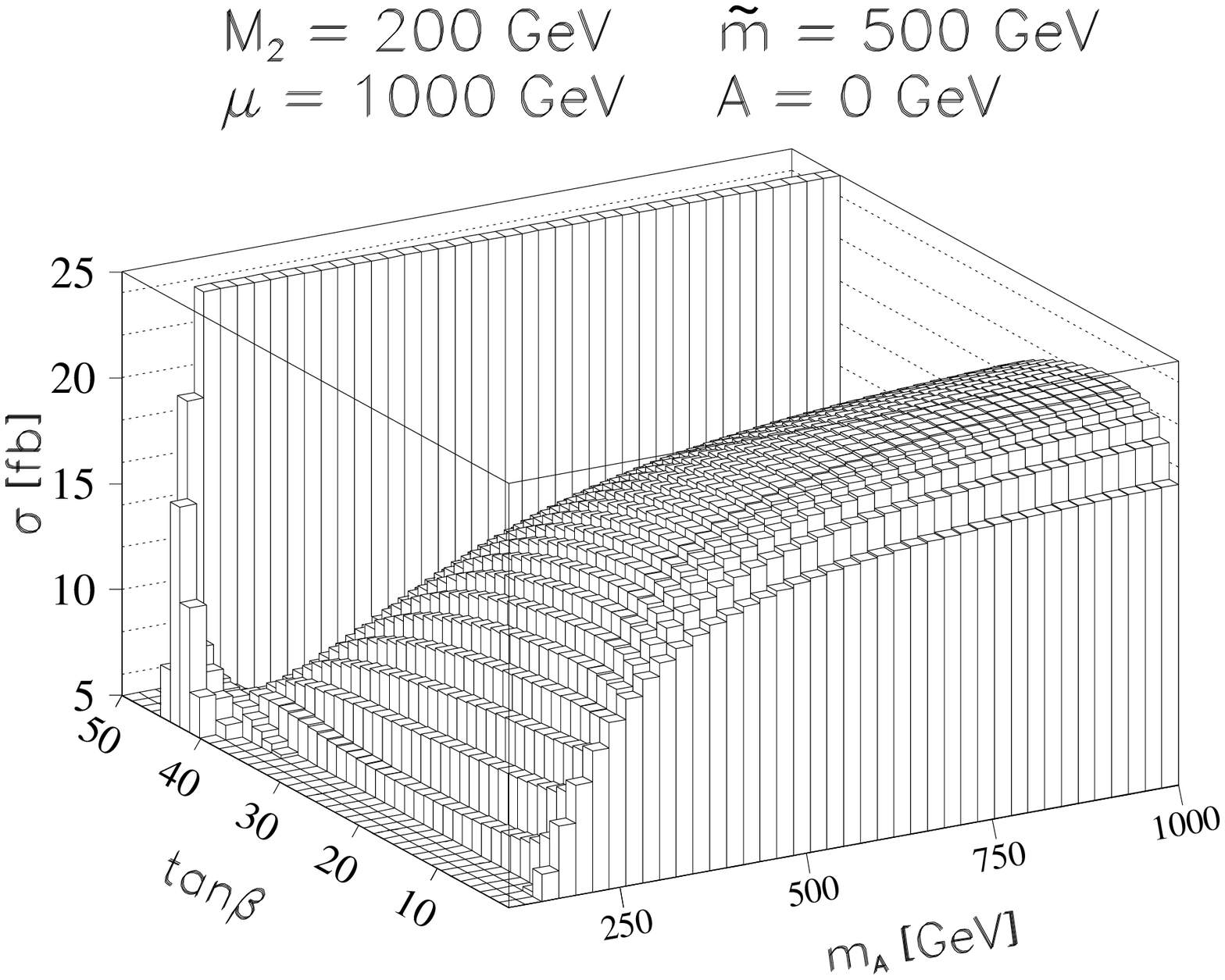}}
 \mbox{\epsfysize=6.0cm\epsffile{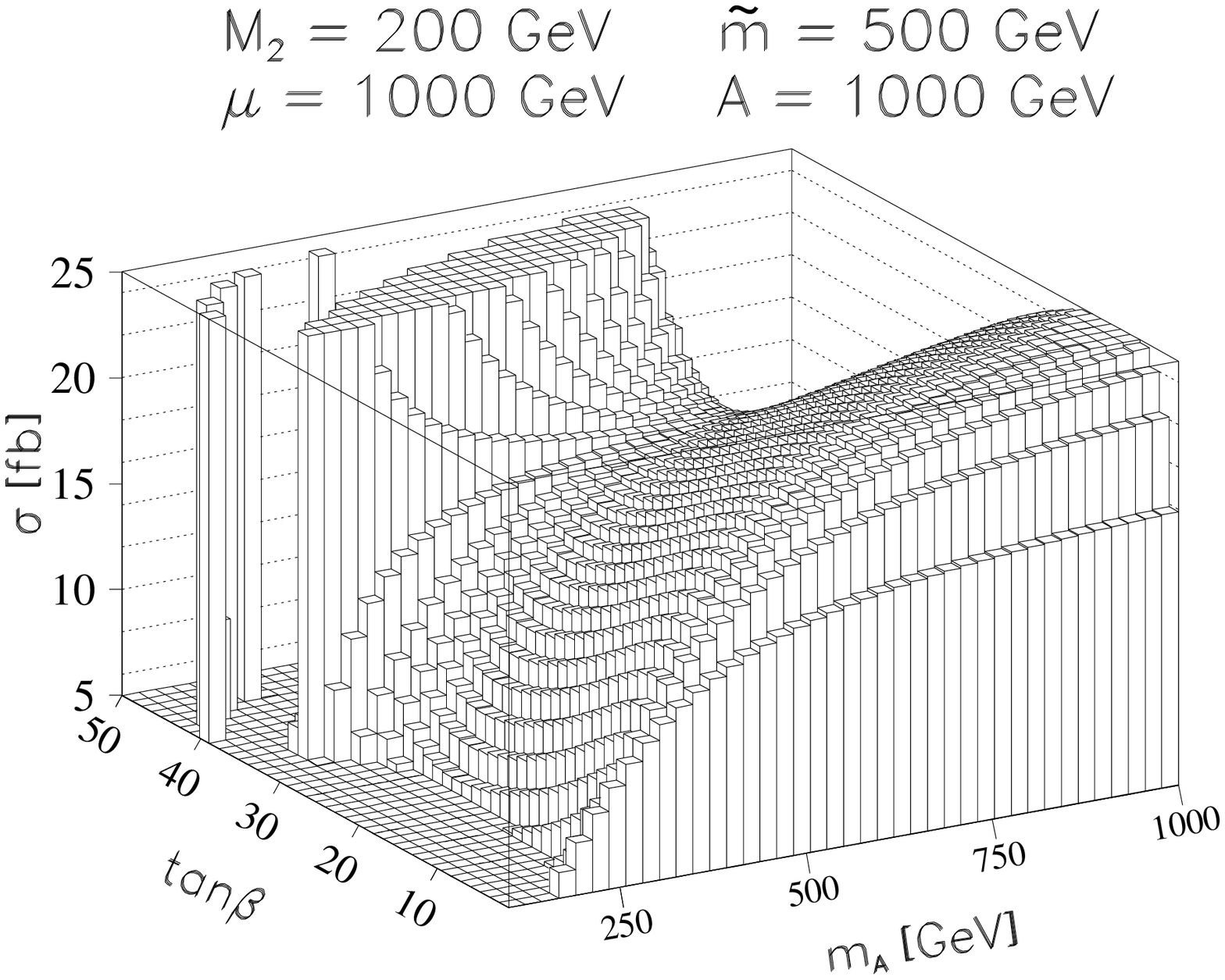}}}
\end{picture}
\begin{capt}
Cross section for $pp\to h^0X\to\gamma\gamma X$ 
as a function of $m_A$ and $\tan\beta$
for $\MNS=200~\GeV$,
$\Ms=500~\GeV$, $\mu=1000~\GeV$, and two values of $A$:
{\em left}: $A=0$, {\em right}: $A=1000~\GeV$.
\end{capt}
\end{center}
\end{figure}
%%%%%%%%%%%%%%%%%%%%%%%%%%%%%%%%%%%%%%%%%%%%%%%%%%%%%%%%%%%%%%%%%%%%%%%%
For large $m_A$, at fixed $\beta$, all Higgs masses, except $m_{h^0}$,
become large, so that $h^0$ decouples. For large $m_{A}$, we
actually have $\sin(\beta-\alpha)\to1$,
which is why the cross section increases and reaches a plateau
for large $m_{A}$.
For these values of $\Ms$ and $\mu$ (fig.~5),
the increase in $A$ leads to larger cross sections over most
of the $m_{A}$--$\tan\beta$ plane.

If we use Higgs masses as given in ref.\ \cite{Carena},
the cross section exhibits less variation with $A$.
But this comparison is incomplete since those masses are not
valid for large $A$ where the squark mass splitting is large.

The $\mu$-dependence of the cross section can for the case of 
$\MNS=200~\GeV$ and $\Ms=500~\GeV$ be described as follows.
At moderate values ($\mu=\pm200~\GeV$), there is not much difference
between the cross section for positive and negative values of $\mu$.
The cross section has a significant dependence on $m_A$, 
being low at $m_A\le\Order(300~\GeV)$, 
then increasing steadily and reaching a plateau with increasing $m_A$.
The dependence on $\tan\beta$ is rather weak.
In \cite{KOP} we show more details in contour plots of the cross section
(\ref{eq:pp-h0-gammagamma}), for $\MNS=200~\GeV$, $\Ms=500~\GeV$
and $\mu$ equal to $\pm500$.

%%%%%%%%%%%%%%%%%%%%%%%%%%%%%%%%%%%%%%%%%%%%%%%%%%%%%%%%%%%%%%%%%%%%%%%%
\begin{figure}[thb]
\begin{center}
\setlength{\unitlength}{1cm}
\begin{picture}(16,5.5)
\put(0.0,-1)
{\mbox{\epsfysize=6.0cm\epsffile{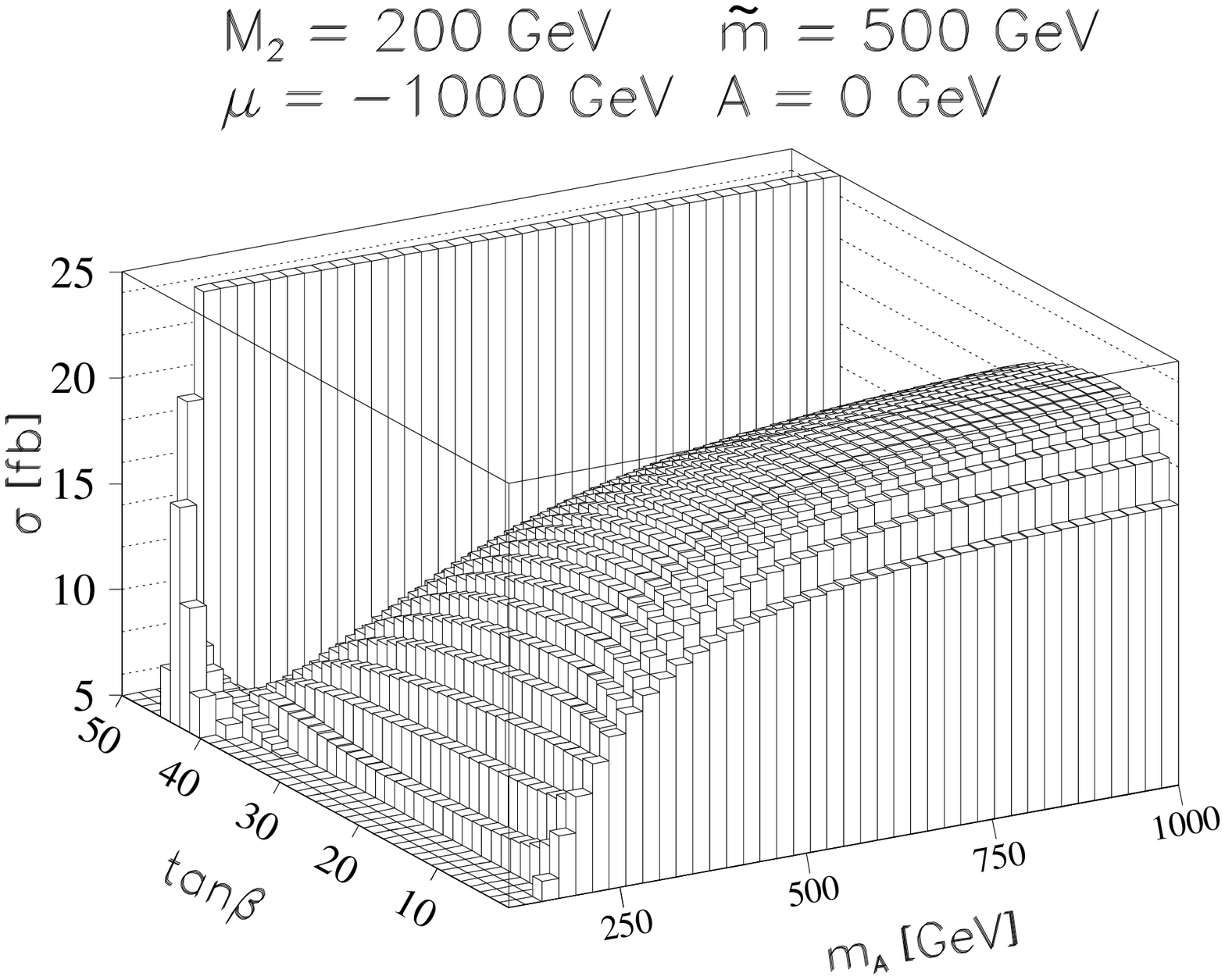}}
 \mbox{\epsfysize=6.0cm\epsffile{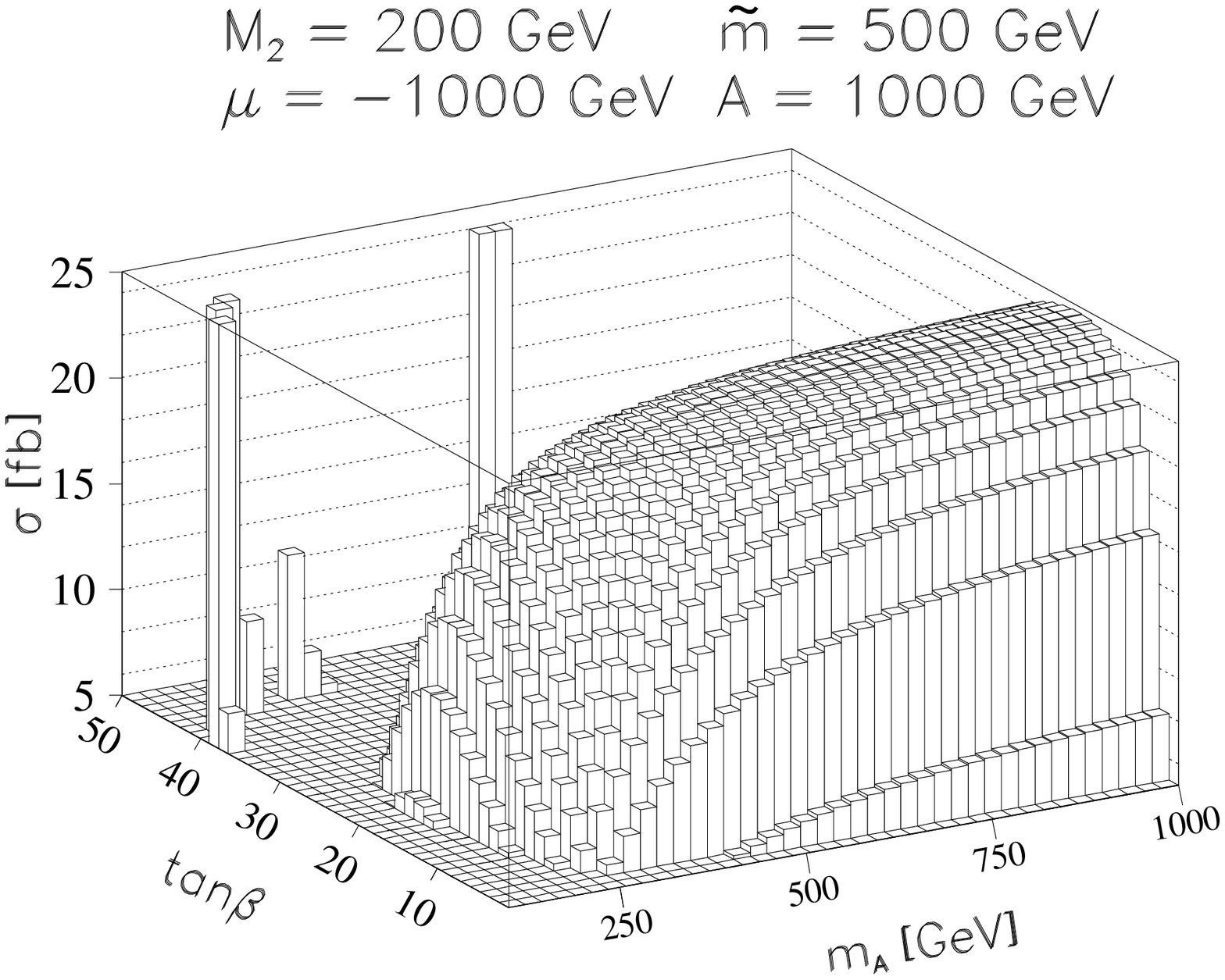}}}
\end{picture}
\begin{capt}
Cross section for $pp\to h^0X\to\gamma\gamma X$
as a function of $m_A$ and $\tan\beta$
for $\MNS=200~\GeV$, $\Ms=500~\GeV$, and $\mu=-1000~\GeV$.
{\em left}: $A=0$, 
{\em right}: $A=1000~\GeV$.
\end{capt}
\end{center}
\end{figure}
%%%%%%%%%%%%%%%%%%%%%%%%%%%%%%%%%%%%%%%%%%%%%%%%%%%%%%%%%%%%%%%%%%%%%%%%

%%%%%%%%%%%%%%%%%%%%%%%%%%%%%%%%%%%%%%%%%%%%%%%%%%%%%%%%%%%%%%%%%%%%%%%%
\begin{figure}[thb]
\begin{center}
\setlength{\unitlength}{1cm}
\begin{picture}(16,5.5)
\put(0.0,-1)
{\mbox{\epsfysize=6.0cm\epsffile{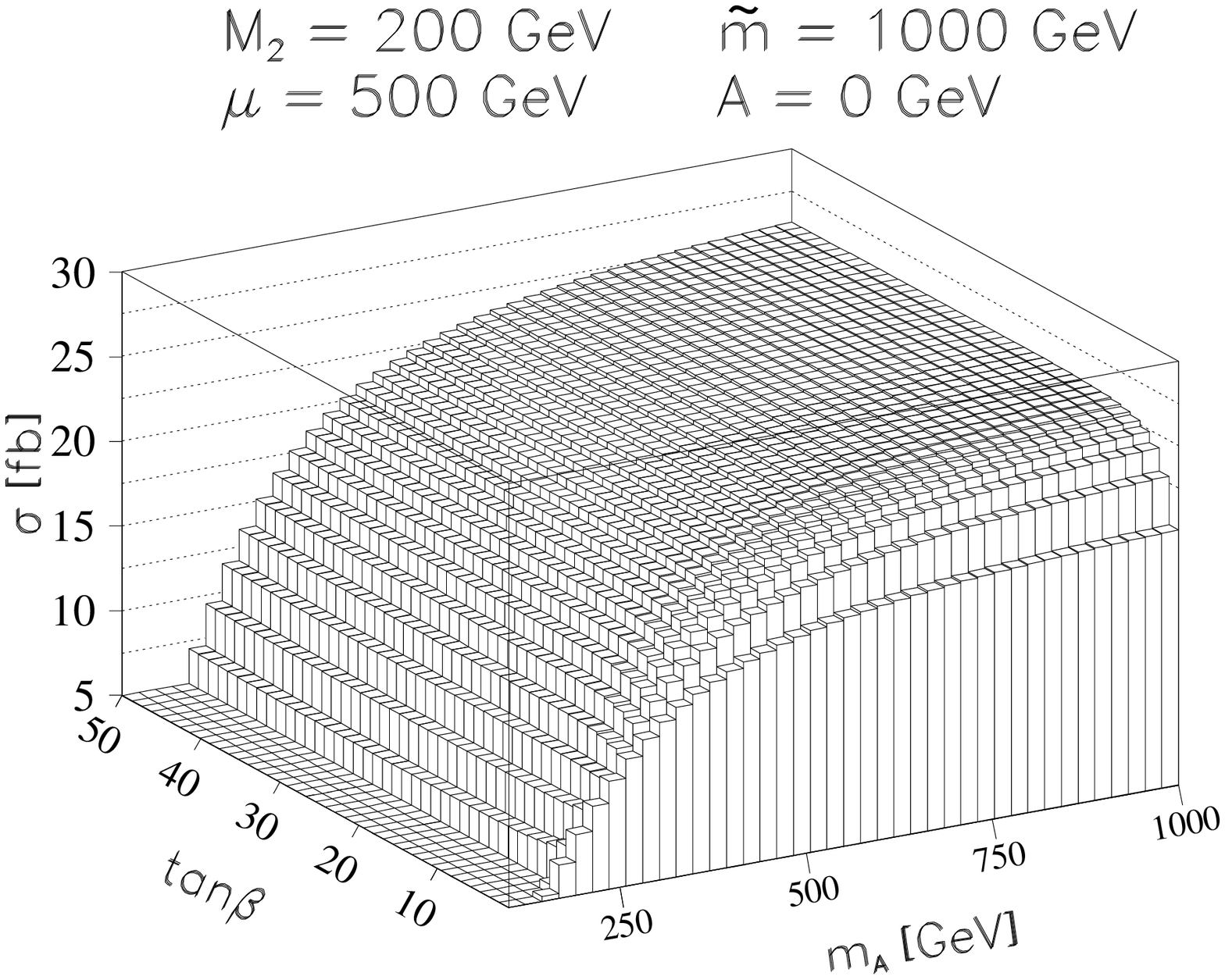}}
 \mbox{\epsfysize=6.0cm\epsffile{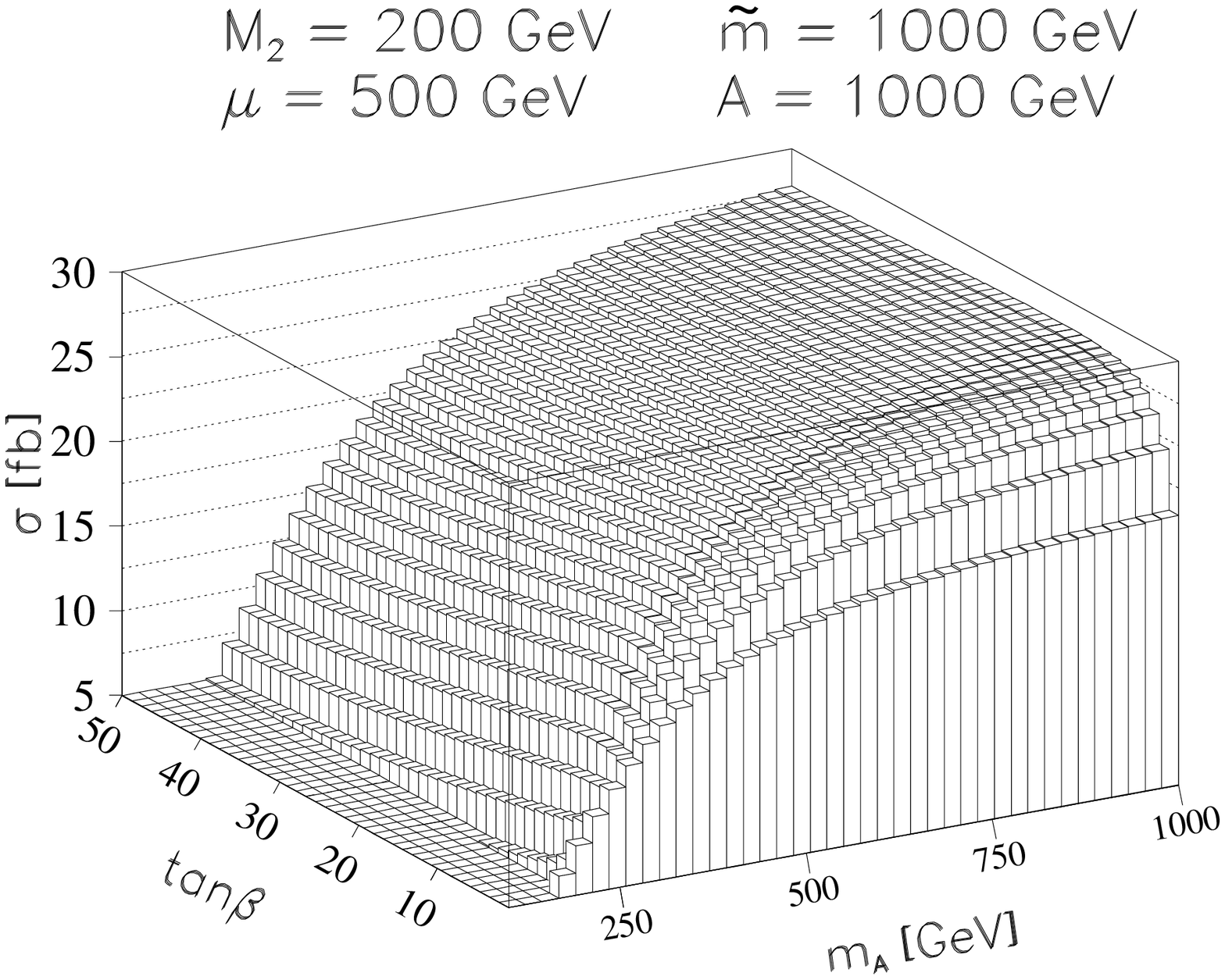}}}
\end{picture}
\begin{capt}
Cross section for $pp\to h^0X\to\gamma\gamma X$
as a function of $m_A$ and $\tan\beta$
for $\MNS=200~\GeV$, $\Ms=1~\TeV$, and $\mu=500~\GeV$.
{\em left}: $A=0$, 
{\em right}: $A=1000~\GeV$.
\end{capt}
\end{center}
\end{figure}
%%%%%%%%%%%%%%%%%%%%%%%%%%%%%%%%%%%%%%%%%%%%%%%%%%%%%%%%%%%%%%%%%%%%%%%%
For increasing values of $|\mu|$ (500~GeV, 1000~GeV)
the change in the cross section is rather complex.
This is illustrated by figures 6--8, and
is basically caused by two phenomena: 
(1) At large values of $|\mu|$ some squarks become too light 
or unphysical, in analogy with the case ($\MNS=200~\GeV$,
$\Ms=150~\GeV$) shown in fig.~1.
Hence, there are regions both at small and large values of $\tan\beta$
where the cross section is not defined.
(2) At large values of $|\mu|$ and large values of $\tan\beta$
(all $m_A$) the Higgs gets very light (due to radiative
corrections, see fig.~2). 

As a consequence of (2), the cross section can get rather high, 
where not forbidden due to (1) above.
However, the two-photon decay rate, which is typically dominated
by the $W$-loop contribution, also depends on the Higgs mass.
In fact, the contribution of the $W$-loop to the decay amplitude
is proportional to \cite{HHG}
\begin{equation}
\label{eq:W-loop}
\frac{\sin(\beta-\alpha)}{\tau} 
\left[2 +3\tau +3\tau(2-\tau)f(\tau) \right],
\end{equation}
with $\tau=(2m_W/m_{h^0})^2$ and $f(\tau)$ the usual `triangle
function'
\begin{equation}
f(\tau)=\cases{\left[\arcsin\left(\frac{1}{\sqrt{\tau}}\right)\right],
\qquad \hspace{18mm} \tau\ge1, \cr
-\left[-\cosh^{-1}\left(\frac{1}{\sqrt{\tau}}\right)+\frac{i\pi}{2}\right]^2,
\qquad \tau<1.}
\end{equation}
In fact, the decrease of this function (\ref{eq:W-loop})
explains the decrease in the cross section multiplied by the branching
ratio, as seen in figs.~6 and 7 for large values of $\tan\beta$
where $m_{h^0}$ becomes small, cf.~fig.~2.
(The coupling factor $\sin(\beta-\alpha)$ remains close to $1$
for practically the whole range of $\tan\beta$, provided $m_A$
is not too small.\footnote{For a discussion of these parameters,
see also ref.~\cite{Djouadi}.})
Finally, the secondary {\it increase} seen in figs.~5 and 6 
at large values of $\tan\beta$ and large $A$ is due to an increase in the
two-photon branching ratio. (The total decay rate falls off faster
at large values of $\tan\beta$ than the two-photon decay rate.)

The dependence of the cross section on $\MNS$ and $\Ms$ is
described in table~1 of ref.~\cite{KOP}.
For small values of $\Ms$ ($\sim150~\GeV$), the possible ranges of
$\tan\beta$, $\mu$ and $A$ become severely restricted, when requiring
physically acceptable squark masses.
There is a significant increase in the cross section as $\Ms$ 
increases from 500~GeV to 1~TeV, to values of the order of 25--30~fb
(see fig.~8). At this large value of $\Ms$, the dependence on $\mu$,
$A$ and $\tan\beta$ becomes weaker.

For small values of $\MNS$ ($\sim50~\GeV$), the allowed range in $\mu$
must be restricted in order to obtain physically acceptable 
chargino masses.
As $\MNS$ {\it increases} beyond 200~GeV, 
there is little further change in the cross section.  

The cross section has a modest dependence on the choice of
gluon distribution function used.
For the plots shown here, we have used the recent GRV Set~3 \cite{GRV}
distributions, which are the default of the PDFLIB.
Other sets lead to variations of the order of 5--10\% \cite{KOP}.
These uncertainties are thus rather insignificant as compared to
the dependence on the mixing parameters $\mu$ and $A$.
%\clearpage
%%%%%%%%%%%%%%%%%%%%%%%%%%%%%%%%%%%%%%%%%%%%%%%%%%%%%%%%%%%%%%%%%%%%%%%%
\section{The heavier $CP$-even Higgs boson $H^0$}
\label{sec:Xsects-H}
%\setcounter{equation}{0}
%%%%%%%%%%%%%%%%%%%%%%%%%%%%%%%%%%%%%%%%%%%%%%%%%%%%%%%%%%%%%%%%%%%%%%%%
Finally, we consider the process
\beq
\label{eq:pp-H0-gammagamma}
pp\to H^0 X \to \gamma\gamma X,
\eeq
which is of interest for small values of $m_A$.

%%%%%%%%%%%%%%%%%%%%%%%%%%%%%%%%%%%%%%%%%%%%%%%%%%%%%%%%%%%%%%%%%%%%%%%%
\begin{figure}[hbt]
\begin{center}
\setlength{\unitlength}{1cm}
\begin{picture}(16,8.5)
\put(3.5,-3)
{\mbox{\epsfysize=12cm\epsffile{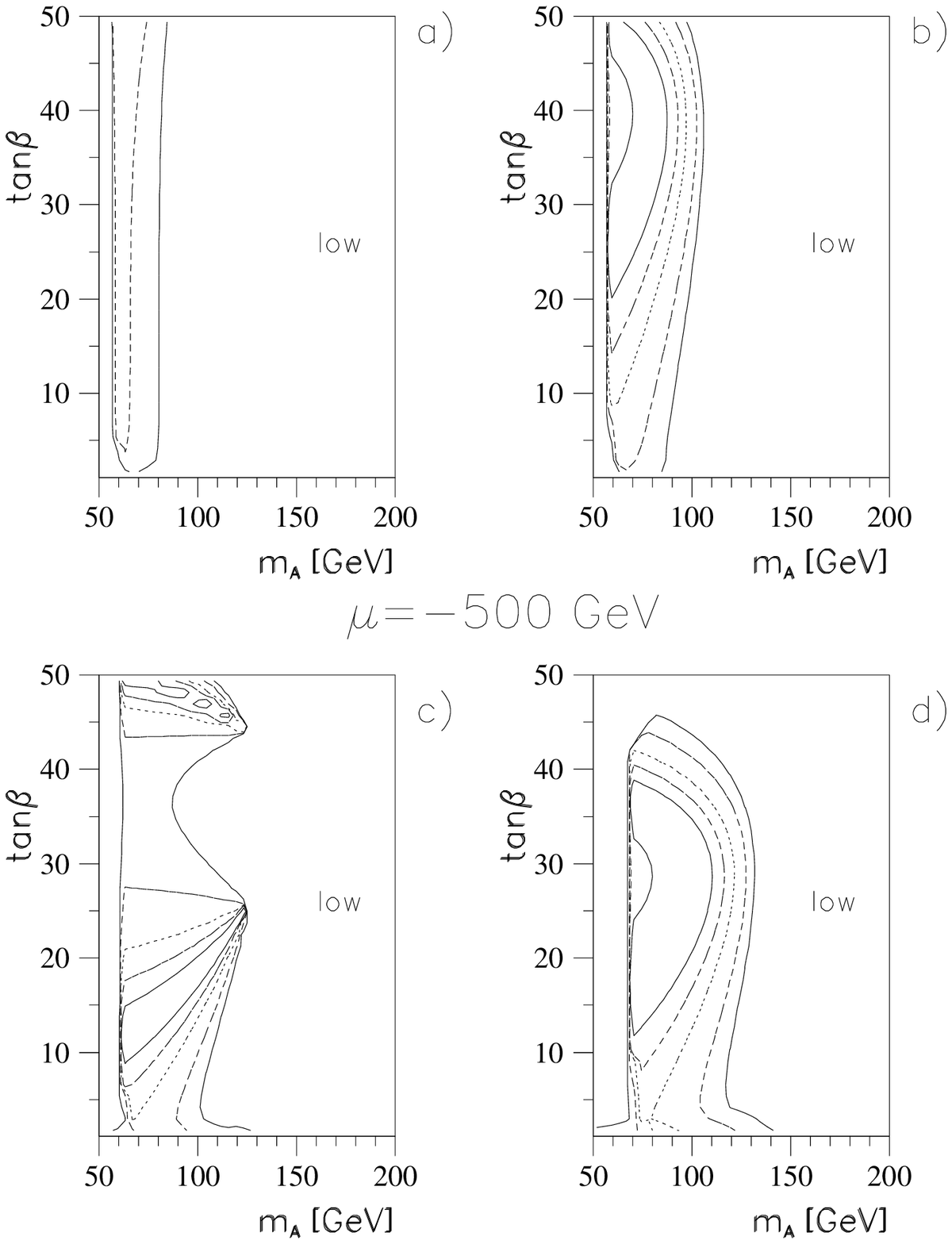}}}
\end{picture}
\vspace*{14mm}
\begin{capt}
Dependence of the $pp\to H^0\to \gamma\gamma$
cross section on $m_A$ and $\tan\beta$
for different values of the trilinear couplings $A$.
Four values of $A$ are considered: a)~$A=0$,
b)~$A=200~\GeV$, c)~$A=500~\GeV$ and d)~$A=1000~\GeV$.
Here $\MNS=200~\GeV$, $\Ms=500~\GeV$, and $\mu=-500~\GeV$.
The contours are at 10~fb (solid), 20~fb, 50~fb, 100~fb and 200~fb.
\end{capt}
\end{center}
\end{figure}
%%%%%%%%%%%%%%%%%%%%%%%%%%%%%%%%%%%%%%%%%%%%%%%%%%%%%%%%%%%%%%%%%%%%%%%%
The two-photon decay of the heavier $CP$-even Higgs boson $H^0$
proceeds dominantly through the $W$ loop, and its amplitude is
proportional to $\cos(\beta-\alpha)$. It is, thus, complementary 
to the decay of the lighter $CP$-even Higgs boson $h^0$, 
whose coupling to a pair of $W$ bosons is proportional to 
$\sin(\beta-\alpha)$. It is significant
only if $m_A$ is small, which means $m_{H^0}$ itself is light, 
since\footnote{In numerical calculations, we use the complete one-loop
radiatively corrected formula for $m_{H^0}$ with non-zero values of
$\mu$, $A_u$ and $A_d$.},
from eqs.\ (\ref{eq:M-matrix}) and (\ref{eq:Deltas}), 
we have (for $\mu = A_u = A_d = 0$)
\begin{equation}
m_{H^0}^2 = m_{A^0}^2 + m_Z^2 - m_{h^0}^2 + \Delta_{22}.
\end{equation}
At small values of $m_A$, the total $H^0$ decay rate is small and thus, 
there can be considerable branching ratio for it to go into two photons. 
As a result, the cross section for the process (\ref{eq:pp-H0-gammagamma})
at small values of $m_A$ can be as large as 200~fb or even more.
%%%%%%%%%%%%%%%%%%%%%%%%%%%%%%%%%%%%%%%%%%%%%%%%%%%%%%%%%%%%%%%%%%%%%%%%
\begin{figure}[thb]
\begin{center}
\setlength{\unitlength}{1cm}
\begin{picture}(16,8.5)
\put(3.5,-3)
{\mbox{\epsfysize=12cm\epsffile{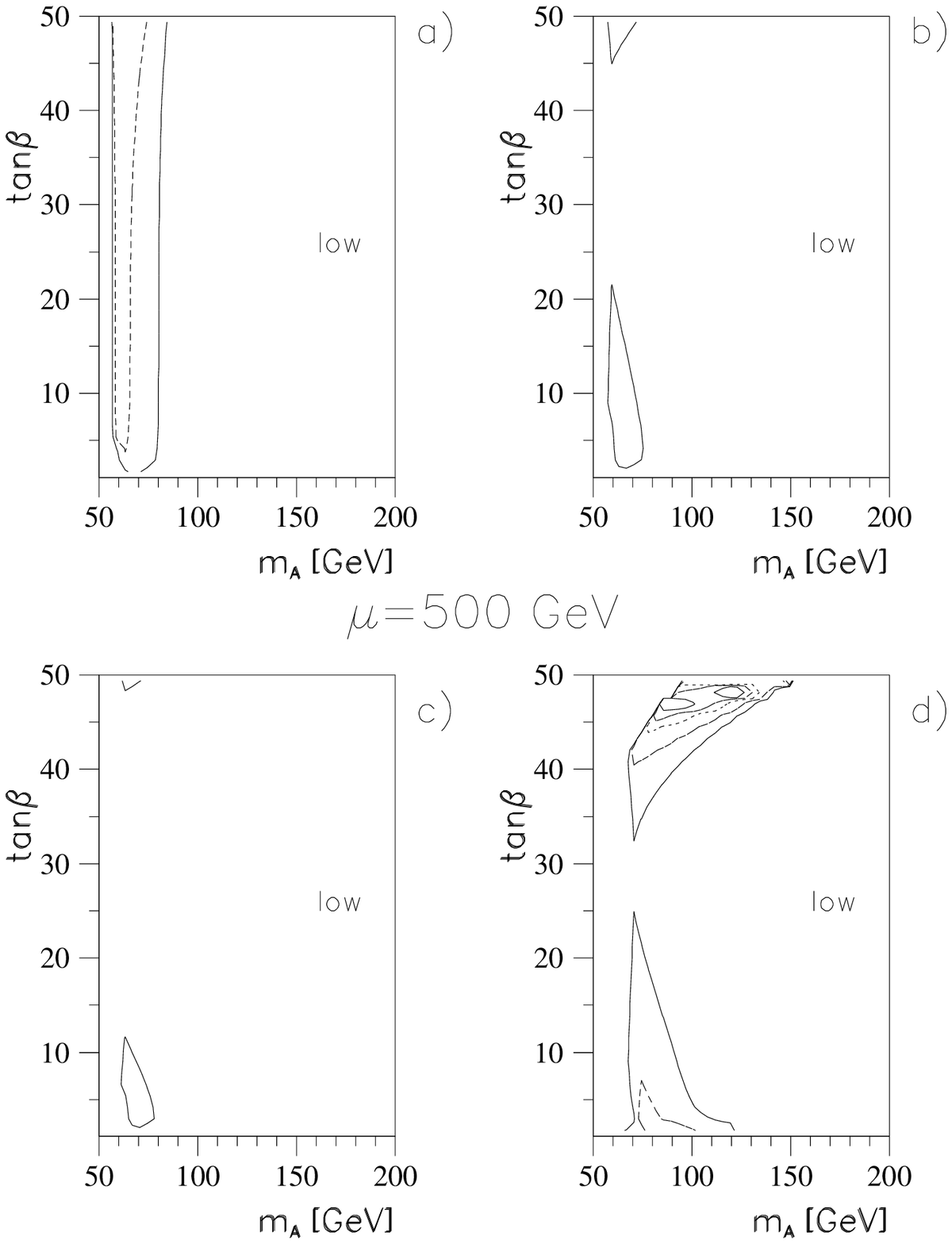}}}
\end{picture}
\vspace*{14mm}
\begin{capt}
Dependence of the $pp\to H^0\to \gamma\gamma$
cross section on $m_A$ and $\tan\beta$
for different values of the trilinear couplings $A$.
As in fig.~9, except that $\mu=500~\GeV$.
\end{capt}
\end{center}
\end{figure}
%%%%%%%%%%%%%%%%%%%%%%%%%%%%%%%%%%%%%%%%%%%%%%%%%%%%%%%%%%%%%%%%%%%%%%%%
In fig.~(9) we show the contour plot of the cross section for the
process (\ref{eq:pp-H0-gammagamma}) for $\mu=-500~\GeV$, 
and for four different values of $A$. For this value of $\mu$, 
there is a strong increase in the cross section with increasing
values of $A$. On the other hand, for positive values of $\mu$, 
increasing the value of $A$ leads to a decrease in the cross section. 
This is shown in fig.~(10), where we plot the cross section for 
$\mu=500~\GeV$. At low values of $m_A$ the Higgs mass $m_{H^0}$ is 
of the order of 110--140~GeV,
and the $h^0$ mass is close to the experimental lower limit.

%%%%%%%%%%%%%%%%%%%%%%%%%%%%%%%%%%%%%%%%%%%%%%%%%%%%%%%%%%%%%%%%%%%%%%%%
\section{Summary and concluding remarks}
\label{sec:conc}
%\setcounter{equation}{0}
%%%%%%%%%%%%%%%%%%%%%%%%%%%%%%%%%%%%%%%%%%%%%%%%%%%%%%%%%%%%%%%%%%%%%%%%
We have discussed in detail the cross section for
the production of $CP$-even Higgs bosons at the LHC, 
in conjunction with their decay to two photons.
Where the parameters lead to a physically acceptable phenomenology,
the cross section multiplied by the two-photon branching ratio
for the lighter $CP$-even Higgs boson is of the order of 20--30~fb.

Similar results have been presented in \cite{Kane}.
Within the context of a SUGRA GUT model, these authors consider
basically a random sample of parameters compatible with
experimental and theoretical constraints.
The cross sections obtained in \cite{Kane} appear to be somewhat
higher than those of \cite{KOP}.

These calculations do not take into account QCD corrections.
Such corrections have been evaluated for the quark-loop contribution,
and lead to enhancements of the cross section of about 50\% \cite{Spira}.
However, in the presence of chiral mixing the squark loops also
contribute significantly. Since the QCD corrections for these
are not available, we have not considered the
higher-order QCD effects here.
One should, of course, keep in mind that they are very important.

There is a modest increase of the cross section with increasing
values of $A$ (i.e., with increasing chiral mixing).
This comes about as the result of two competing effects:
with increasing $A$, the Higgs boson becomes more heavy, 
leading to lower production cross sections.
This is however offset by a corresponding increase 
in the two-photon decay rate.

\medskip
It is a pleasure to thank the Organizers of the Zvenigorod Workshop,
in particular Professor V. Savrin,
for creating a very stimulating and pleasant atmosphere during the meeting.
This research has been supported by the Research Council of Norway.
PNP would like to thank Alexander von Humboldt-Stiftung 
and Prof.~H.J.W.~M\"uller-Kirsten
for support while this paper was written.
The work of PNP is supported by the Department of Science
and Technology under project No. SP/S2/K-17/94.
%%%%%%%%%%%%%%%%%%%%%%%%%%%%%%%%%%%%%%%%%%%%%%%%%%%%%%%%%%%%%%%%%%%%%%%%
                
%%%%%%%%%%%%%%%%%%%%%%%%%%%%%%%%%%%%%%%%%%%%%%%%%%%%%%%%%%%%%%%%%%%%%%%%

\end{document}